\documentclass[conference]{IEEEtran}
\IEEEoverridecommandlockouts
\usepackage{cite}
\usepackage{amsmath,amssymb,amsfonts}
\usepackage{algorithmic}
\usepackage{graphicx}
\usepackage{textcomp}
\usepackage{xcolor}
\def\BibTeX{{\rm B\kern-.05em{\sc i\kern-.025em b}\kern-.08em
    T\kern-.1667em\lower.7ex\hbox{E}\kern-.125emX}}
\begin{document}

\title{TPM: A GPS-based Trajectory Pattern Mining System}

\author{
	\IEEEauthorblockN{Yang Cao, Jingling Yuan, Song Xiao, Qing Xie} 
	
	\IEEEauthorblockA{School of Computer Sci \& Technol, Wuhan University of Technology, Wuhan, China\\
	c\_young514@163.com, yuanjingling@126.com, 664621491@qq.com, felixxq@whut.edu.cn} 
}

\maketitle

\begin{abstract}
With the development of big data and artificial intelligence, the technology of urban computing becomes more mature and widely used. In urban computing, using GPS-based trajectory data to discover urban dense areas, extract similar urban trajectories, predict urban traffic, and solve traffic congestion problems are all important issues. This paper presents a GPS-based trajectory pattern mining system called TPM. Firstly, the TPM can mine urban dense areas via clustering the spatial-temporal data, and automatically generate trajectories after the timing trajectory identification. Mainly, we propose a method for trajectory similarity matching, and similar trajectories can be extracted via the trajectory similarity matching in this system. The TPM can be applied to the trajectory system equipped with the GPS device, such as the vehicle trajectory, the bicycle trajectory, the electronic bracelet trajectory, etc., to provide services for traffic navigation and journey recommendation. Meantime, the system can provide support in the decision for urban resource allocation, urban functional region identification, traffic congestion and so on.
\end{abstract}

\begin{IEEEkeywords}
urban computing, spatio-temporal data, similar trajectory, K-means clustering
\end{IEEEkeywords}

\section{Introduction}
Urbanization’s rapid progress has modernized many people’s lives but also brought many problems such as population, traffic congestion, pollution, energy consumption, and complex social management. With the development of artificial intelligence and big data technologies, Artificial intelligence-driven urban computing or smart cities have also grown rapidly. Since the maturity of urban sensing technology and computing environment, a lot of spatio-temporal data, such as vehicle trajectory data, traffic snapshots, road network data, etc., are generated in the city. It is especially important to use these spatio-temporal data and AI technology to drive smart cities. Advanced cities have introduced urban transportation convenience programs such as shared bicycles or shared taxis, public transportation vehicles, and station navigation. Therefore, we can use the spatio-temporal data to analyze and model, and mine more valuable information on trajectory data. And then, we can provide decision support for urban life, planning and construction from the perspective of spatio-temporal data mining.

However, in urban computing, the existing studies are usually dependent on the road network, and the bounding of trajectory and road network requires a lot of data preprocessing. In addition, in some scenarios, there is no complete road networks, such as underdeveloped area, sea navigation, air aviation, etc. Therefore, it is very necessary to develop a trajectory pattern mining system that does not depend on the road network. 

\begin{figure*}[htbp]
\centerline{\includegraphics[scale=0.48]{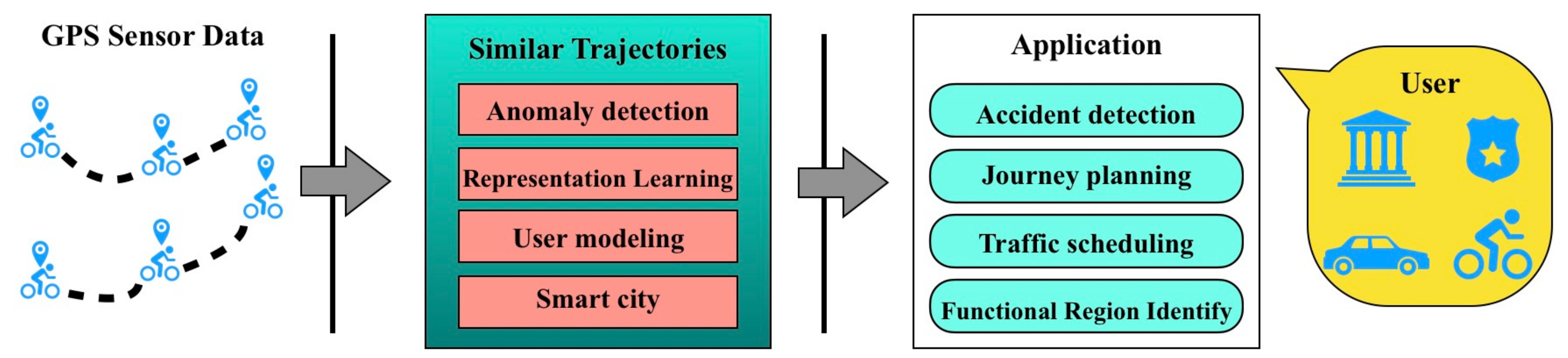}}
\caption{Application background of the TPM}
\label{fig1}
\end{figure*}

Considering the problems above, we propose to design a GPS-based trajectory pattern mining system called TPM. In the TPM, we can automatically get urban dense areas and GPS trajectories. Importantly, we can mine similar trajectory patterns via the trajectory data similarity matching. These are not required to bind the road network. We analyze the GPS data, first extracts the travel trajectory, and then mines similar pattern, anomaly detection, representation learning, etc. according to the trajectory features, for accident detection, journey planning, traffic scheduling, functional region identity provide support, and its application background shown in Figure \ref{fig1}.

In the TMP, we only apply the GPS-based trajectory instead of the road network. We use K-means to cluster the points of the trajectory into virtual regions, each of which corresponds to a place in the map, and then match these virtual regions in order to obtain similar trajectory patterns, thus this system is not restricted by the road network and has better robustness.

\section{Related Work}
In recent years, more and more scholars have paid attention to the study of spatio-temporal data in cities. Zheng \cite{b1}\cite{b2} et al. mentioned that urban computing connects urban sensing, data management, data analytics, and service providing into a recurrent process for an unobtrusive and continuous improvement of people’s lives, city operation systems, and the environment. In this field, the application of the GPS-based trajectory is more important. Following a road map from the derivation of trajectory data, to trajectory data preprocessing, to trajectory data management, and to a variety of mining tasks (such as trajectory pattern mining, outlier detection, and trajectory classification), they explore the connections, correlations, and differences among these existing techniques. Giannotti \cite{b3} et al. developed an extension of the sequential pattern mining paradigm that analyzes the trajectories of moving objects. They introduce trajectory patterns as concise descriptions of frequent behaviors, in terms of both space and time.

There are some studies of trajectory pattern mining. The work in \cite{b4} proposes a framework titled that discovers regions of different functions in a city using both human mobility among regions and points of interests (POIs) located in a region. They infer the functions of each region using a topic-based inference model, which regards a region as a document, a function as a topic, categories of POIs as metadata, and human mobility patterns as words. The work in \cite{b5} is planning bike lanes based on sharing-bikes' trajectories. They proposed a data-driven approach to develop bike lane construction plans based on large-scale real-world bike trajectory data and considered key realistic constraints: budget limitations, construction convenience, and bike lane utilization. Two main components are employed in the proposed illegal parking detection system \cite{b6}: one is that trajectory pre-processing, which filters outlier GPS points, performs map-matching and builds trajectory indexes; and the other is that illegal parking detection, which models the normal trajectories, extracts features from the evaluation trajectories and utilizes a distribution test-based method to discover the illegal parking events. And the work in \cite{b7} propose and study the problem of trajectory-driven influential billboard placement. One core challenge is to identify and reduce the overlap of the influence from different billboards to the same trajectories, while keeping the budget constraint into consideration.

Some scholars study trajectory search and trajectory prediction using GPS-based trajectory data. Bao \cite{b8} et al. designed a holistic cloud-based trajectory data management system on Microsoft Azure to bridge the gap between trajectory data and urban applications. The system can efficiently store, index, and query large trajectory data with three functions: trajectory ID-temporal query, trajectory spatio-temporal query, and trajectory map-matching. Wang \cite{b9} et al. study a new kind of query—Reverse k Nearest Neighbor Search over Trajectories, which can be used for route planning and capacity estimation. DITA \cite{b10} exhibits some unique features: trajectory similarity search, join operations and so on. Torch \cite{b11} is able to efficiently process two types of typical queries (similarity search and Boolean search), and there is a new similarity function LORS in Torch to measure the similarity in a more effective and efficient manner.

In addition, some scholars have some deep learning related research. The work in \cite{b12}\cite{b13}\cite{b14} studies subspace clustering for multi-view data while keeping individual views well encapsulated and multiview spectral clustering via structured low-rank matrix factorization. The work in\cite{b15}\cite{b16}\cite{b17}  observes that the same landmarks provided by different users over social media community may convey different geometry information depending on the viewpoints and angles, and may subsequently yield very different results. And then, they propose a novel framework, namely multi-query expansions, to retrieve semantically robust landmarks by two steps. The work in \cite{b18} and \cite{b19}  employs adversarial training scheme to lean a couple of hash functions enabling translation between modalities while assuming the underlying semantic relationship. To induce the hash codes with semantics to the input-output pair, cycle consistency loss is further proposed upon the adversarial training to strengthen the correlations between inputs and corresponding outputs.

\section{System Overview}
\begin{figure}[htbp]
	\centerline{\includegraphics[scale=0.38]{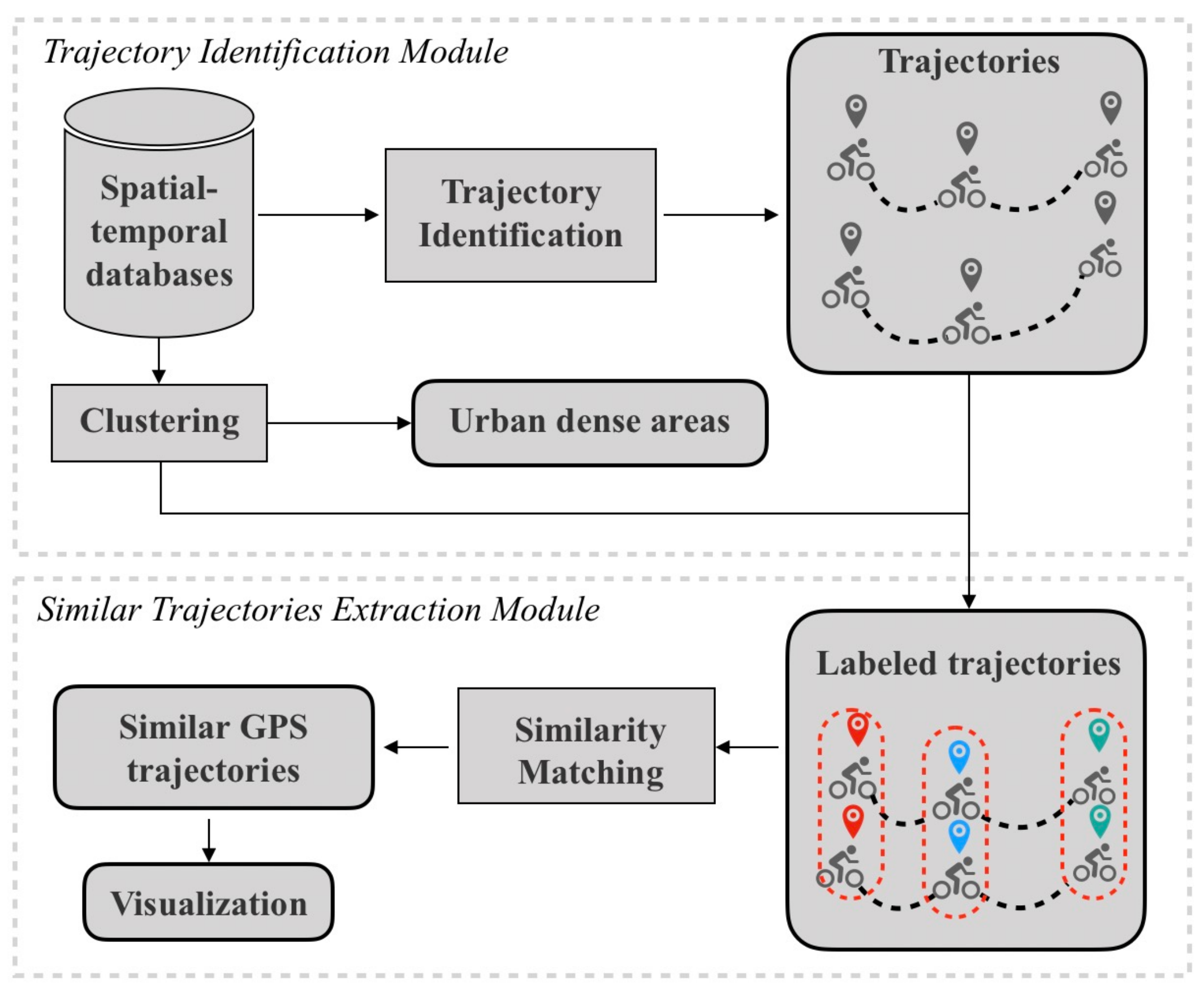}}
	\caption{System overview}
	\label{fig2}
\end{figure}

The TPM system is designed to mine GPS-based trajectory patterns, including obtaining urban dense areas, automatically generating trajectories and extracting similar trajectory patterns in traffic. Figure \ref{fig2} shows the overview of the TPM system, which is divided into two parts. The first part is the trajectory identification module, which mainly preprocesses spatio-temporal data and generates trajectories. In this module, the system also mines urban dense areas via clustering algorithm. This is a basic module that performs preprocessing for the second module in addition to implementing its functions. The second part is the similar trajectories extraction module. In this part, the system labels the coordinates in the trajectory according to the clustering result of the previous module, and then obtains similar trajectories according to the similarity matching rule, and then visualizes them.

\subsubsection{Trajectory Identification Module}
This module is the basic module in the TPM system. It contains two components: trajectory identification and clustering. The trajectory identification component mainly calculates the maximum time interval of the trajectory through massive data analysis. Specifically, each collected coordinate of the GPS device is sorted according to the timestamp sequence, and if the time interval is greater than a certain value, and the number of trajectories does not change significantly, then the time interval value is the basis for the trajectory identification we need. In this way, the system can divide the data of the same GPS device into multiply journeys (trajectories). The clustering component mainly clusters spatio-temporal data according to spatial distribution to obtain urban dense areas, and labels the trajectories for the next module.

\subsubsection{Similar Trajectories Extraction Module}
This module is the core of the TPM system. It mainly makes similarity matching by using the labeled trajectories. In similarity matching, multiple trajectories with the same starting label and the same end label are grouped into the same group. In each group, all co- ordinate labels in the trajectory with less coordinates are defined as set A, and all coordinate labels in the trajectory with more coordinates are defined as set B. If A is included in B, the TPM system determines that they are similar trajectories, that is, similar patterns. And then, the system automatically extracts similar trajectories.

By preprocessing the spatio-temporal data, trajectory identification, clustering, and trajectory similarity matching, the TPM system can automatically generate trajectories, obtain urban dense areas, and extract similar trajectories.

\section{Principle}
\subsection{Trajectory Identification and Preprocessing}
\begin{figure}[htbp]
\centerline{\includegraphics[scale=0.2]{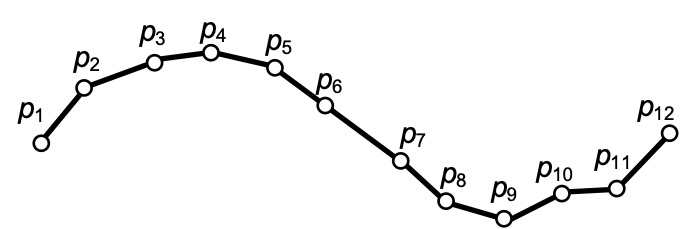}}
\caption{Formation of the trajectory}
\label{fig3}
\end{figure}

Spatio-temporal data has a time and space dimension. The trajectory pattern is a trajectory generated by moving objects in geospatial space, usually represented by a series of chronologically arranged points, for example, ${p\mathop{{}}\nolimits_{{1}} \to p\mathop{{}}\nolimits_{{2}} \to ... \to p\mathop{{}}\nolimits_{{n}}}$, as shown in Figure \ref{fig3}. Each of these points consists of a set of geospatial coordinates and a timestamp, such as $p= ( x,y,t ) $.

In trajectory identification, the coordinate set and time stamp of geospatial are mainly studied. In this paper, we use the shared bicycle dataset, and the data fields used are id, timestamp, and location. Id shares the unique identifier of the bicycle, timestamp uniquely identifies the time of a moment, location contains latitude and longitude information.

In this paper, the data of each id of the collection point is sorted according to the timestamp sequence. The steps for trajectory identification are as follows:
\begin{itemize}
	\item[1)] Divide data by id
	\item[2)] Sort data by timestamp in each id
	\item[3)] Traverse each GPS point of each id
	\item[4)] If time interval less than $\theta$ min: trajectory append point
	\item[5)] Else: id append trajectory
\end{itemize}

\begin{figure}[htbp]
	\centerline{\includegraphics[scale=0.16]{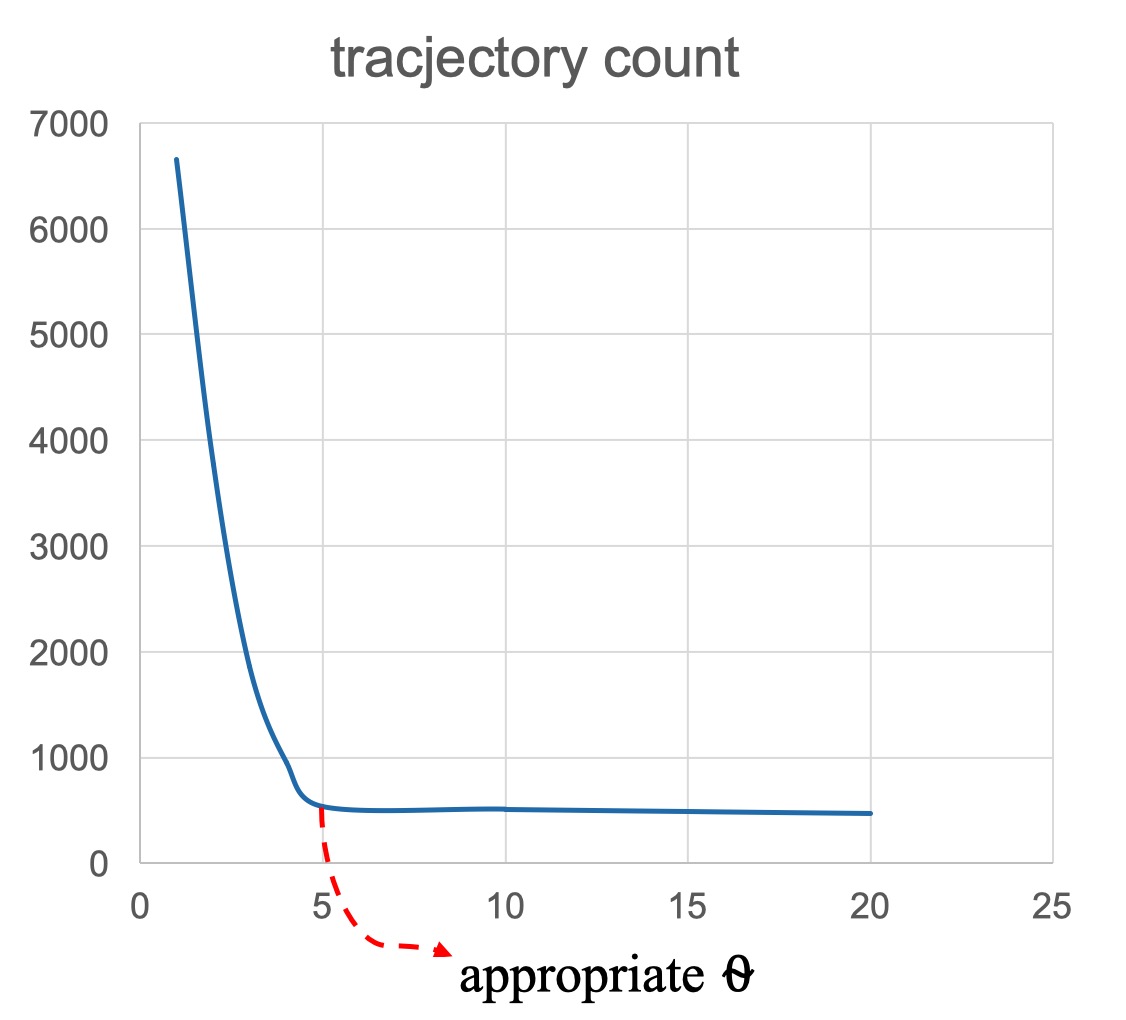}}
	\caption{Appropriate time interval $\theta$}
	\label{fig4}
\end{figure}

We have to find an appropriate time interval  $\theta$ . When the value of  $\theta$  changes, the count of trajectory does not change significantly, and that is the appropriate  $\theta$. That is, we have to get the appropriate  according to the formula $max(|f(\theta)''|)$, as shown in Figure \ref{fig4}.

\begin{figure*}[htbp]
	\centerline{\includegraphics[scale=0.3]{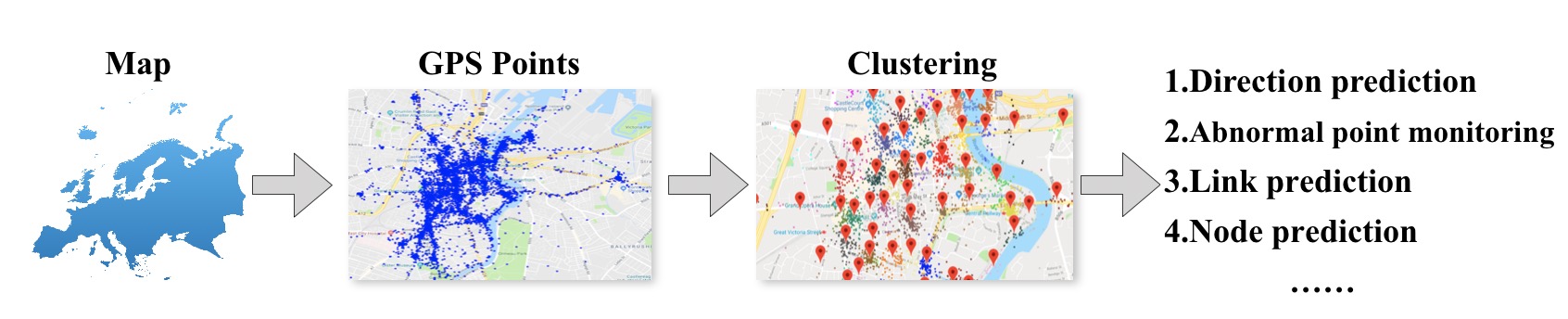}}
	\caption{Clustering diagram}
	\label{fig5}
\end{figure*}

According to the above method for trajectory identification, we can get some rough trajectories. Then calculate the distance of the trajectory according to the latitude and longitude distance formula (3), the formulas are as follows:

	\begin{equation}
	dis1= \sin ^2\Big(\frac{latA-latB}{2}\Big)
		\end{equation}
		\begin{equation}
	dis2=\sin ^2\Big(\frac{lngA-lngB}{2}\Big)
	\end{equation}
  	 \begin{small}
		\begin{equation}
	distance = 2R*arcsin\Big (\sqrt{dis1+\cos(latA)*\cos(latB)*dis2}\Big)
  	    \end{equation}  
      \end{small}

Where $R$ is the radius of the earth, $latA$ and $latB$ are the latitudes of the two points $A$ and $B$, $lngA$ and $lngB$ are the longitudes of the two points $A$ and $B$, and the longitude difference is a small angle.

Spatial trajectories are never perfectly accurate, due to sensor noise and other factors, such as receiving poor positioning signals in urban canyons. Sometimes, the error is acceptable (e.g., a few GPS points of a vehicle fall out of the road the vehicle was actually driven), which can be fixed by map-matching algorithms. In other situations, the error of a noise point is too big (e.g., several hundred meters away from its true location) to derive useful information, such as travel speed. So, we need to filter such noise points from trajectories before starting a mining task. For ease of analysis, the noise filtering in this paper are excluded as following:

\begin{itemize}
\item The trajectory has only a single data point; 
\item There is a data point in the trajectory where the distance is very distant; 
\item The length of the trajectory is too short, it is not suitable to mine similar pattern.
\end{itemize}

\subsection{Clustering of Spatio-temporal Data}
The process of dividing a collection of physical or abstract objects into multiple classes of similar objects is called clustering. A cluster generated by clustering is a collection of data objects that are similar to objects in the same cluster and different from objects in other clusters. There are a large number of classification problems in the natural sciences and social sciences. Cluster analysis, also known as group analysis, is a statistical analysis method for studying classification problems. Cluster analysis originates from taxonomy, but clustering is not equal to classification. The difference between clustering and classification is that the classes required for clustering are unknown. The cluster analysis content is very rich, including systematic clustering method, ordered sample clustering method, dynamic clustering method, fuzzy clustering method, graph theory clustering method, cluster forecasting method and so on.
The clustering algorithm is a relatively mature data mining algorithm. In this paper, the K-means++ algorithm improved from the K-means is used to cluster all the coordinates of the trajectory into k clusters, and the coordinates are k labels. The spatial distribution of these clusters can be regarded as k regions. Therefore, we can perform directional prediction of trajectory data, abnormal point monitoring, link prediction, node prediction, etc. based on the results of clustering, as shown in Figure \ref{fig5}.

The K-means++ algorithm is an improvement on the choice of the initial point, and the other steps are the same as K-means. The basic idea of initial centroid selection is that the distance between the initial cluster centers is as far as possible. This is also a feature required for spatio-temporal data clustering. The K-means++ clustering for spatio-temporal data algorithm process is as follows:

\begin{itemize}
	\item[1)] Randomly select k coordinate points as the initial centroid (K-means++ initialization as far as possible);
	\item[2)] When the centroid of the cluster has changed, calculate the distance between each data and each centroid in the data set;
	\item[3)] Find the nearest cluster and assign it to the update;
	\item[4)] For each cluster, calculate the mean of all points in the cluster, and use the mean as the centroid, repeat the above steps until the centroid of the cluster tends to be stable and the clustering is completed.
\end{itemize}

\subsection{Similarity Matching Rule}
Moving together patterns \cite{b1} is to discover a group of objects that move together for a certain time period, such as flock, convoy, traveling companion and gathering. These patterns can help the study of species’ migration, military surveillance, and traffic event detection, and so on. These patterns can be differentiated between each other based on the following factors: the shape or density of a group, the number of objects in a group, and the duration of a pattern.

Specifically, a flock is a group of objects that travel together within a disk of some user-specified size for at least k consecutive time stamps. A major concern with flock is the predefined circular shape, which may not well describe the shape of a group in reality, and therefore may result in the lossy-flock problem. To avoid rigid restrictions on the size and shape of a moving group, the convoy is proposed to capture generic trajectory pattern of any shape by employing the density-based clustering. Instead of using a disk, a convoy requires a group of objects to be density connected during k consecutive time points.

Based on moving together patterns, we explore a way to find similar trajectory patterns. We analyze the GPS data, and propose similarity matching rule in this paper. And then we can do similar pattern, anomaly detection, representation learning, etc. according to the trajectory features, for accident detection, journey planning, traffic scheduling, functional region identity provide support.

\begin{figure}[htbp]
	\centerline{\includegraphics[scale=0.26]{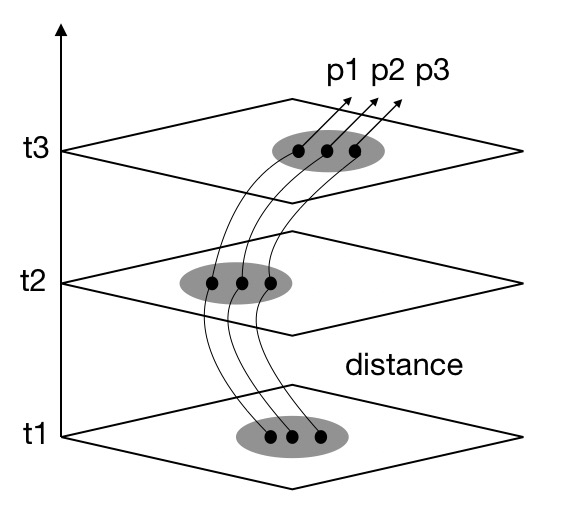}}
	\caption{Similar trajectory model diagram}
	\label{fig6}
\end{figure}

The similarity matching rule is mainly based on the position similarity and sequence of the trajectory. For the similarity of positions, we determine according to the method of clustering. For sequence, we process the trajectory data based on the characteristics of the time series. In addition, we limit the distance of the trajectory to make it close. The similar trajectory model diagram is shown in Figure \ref{fig6}.

Where, $t1$, $t2$, $t3$, etc. is a time series, gray ovals represent similar regions obtained by clustering, and $p1$, $p2$, $p3$ are a group of similar trajectories.

According to the above analysis, the main steps of the similarity matching rule are as follows: 
\begin{itemize}
\item[1)] According to the clustering result, mark the GPS point of the trajectory as a label $n$  $(n=0, 1, ..., k-1)$, and remove the continuous repeating label of the same trajectory;

\item[2)] The trajectory with the same first and last tags is divided into the same set $G$, the maximum difference $d$ of the trajectory distance in $G$ is $d<\varepsilon$, otherwise the trajectory data is cleaned;

\item[3)] In the same set $G$, the label set of the trajectory with the most labels is $A$, the other trajectory label sets are $B1, 2, …, i$, and if $A$ contains $B$, it is a similar trajectory pattern.
\end{itemize}

\section{Experimental Result}
\subsection{Experimental Environment}
We set up the Python runtime platform to configure the Anaconda scientific computing environment. Anaconda is an open source Python distribution that includes more than 180 scientific packages such as Conda and Python and their dependencies. Therefore, we can directly use the toolkits such as numpy and pandas to perform preprocessing and analysis on spatiotemporal data. In addition, we also loaded a machine learning toolkit like scikit-learn. Using this toolkit, we clustered the data and further analysis. As for visualization, we use the Google map API to visualize our spatio-temporal trajectories, urban dense areas, and similar patterns.

\subsection{Experimental Data}
\begin{figure}[htbp]
	\centerline{\includegraphics[scale=0.16]{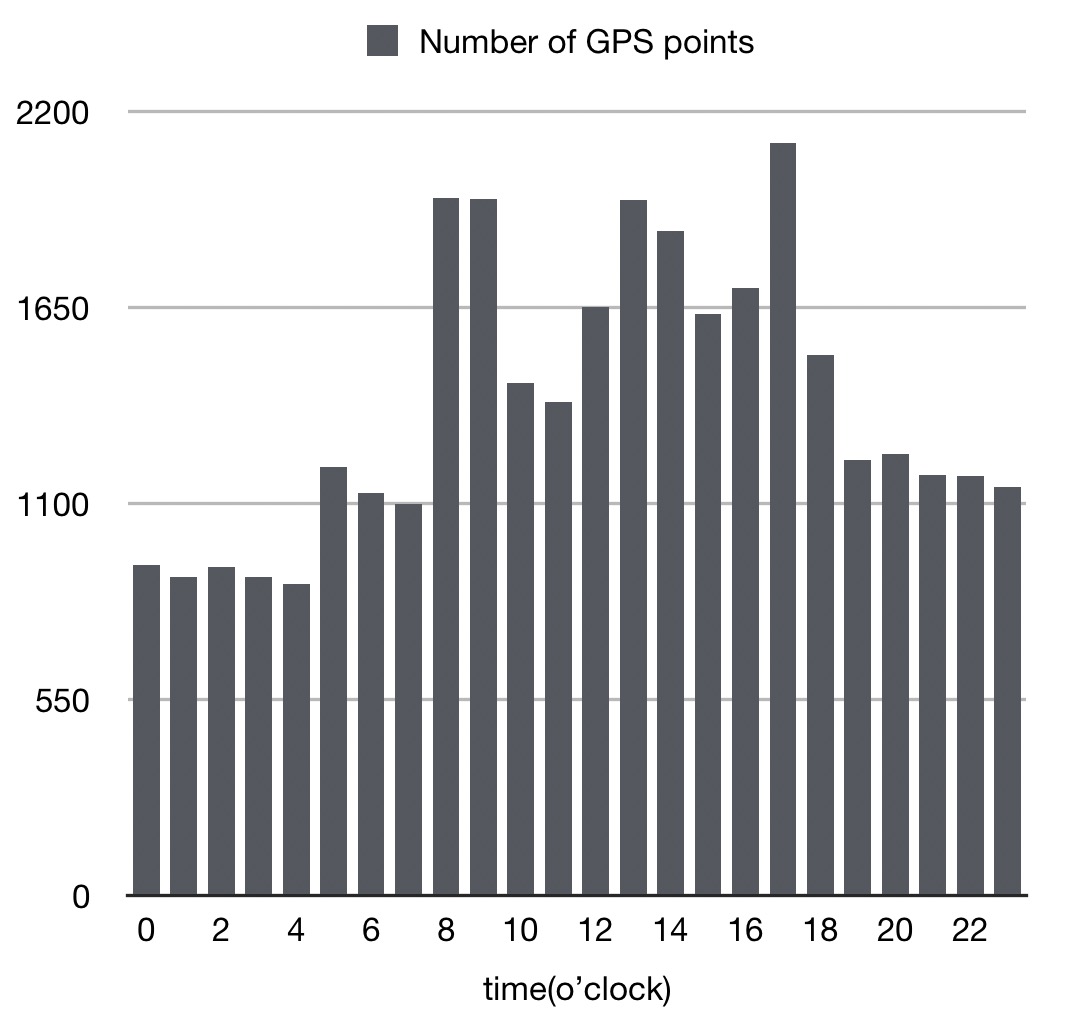}}
	\caption{Time-points diagram}
	\label{fig7}
\end{figure}

This paper uses a city's traffic movement dataset, where the data format is data id, data longitude, data latitude and some additional information that is not used. We analyze the time and space dimensions of these spatiotemporal data. In terms of time, we count all the data and find some rules that are consistent with our daily life. For example, the frequency of the trajectory is relatively highest during a specific time period, such as, from 8:00 am to 9:00 pm, from 1:00 pm to 2:00 pm, and 5 pm, which corresponds to the morning peak, lunch break and evening peak in daily life, as shown in Figure \ref{fig7}. In terms of space, we can find that the GPS points of these trajectories are mostly concentrated in subways, schools, restaurants, entertainment venues and other areas.

\subsection{Experimental Result Demonstration}

\begin{figure*}[htbp]
	\begin{minipage}{0.3\linewidth}
		\centerline{\includegraphics[width=2in]{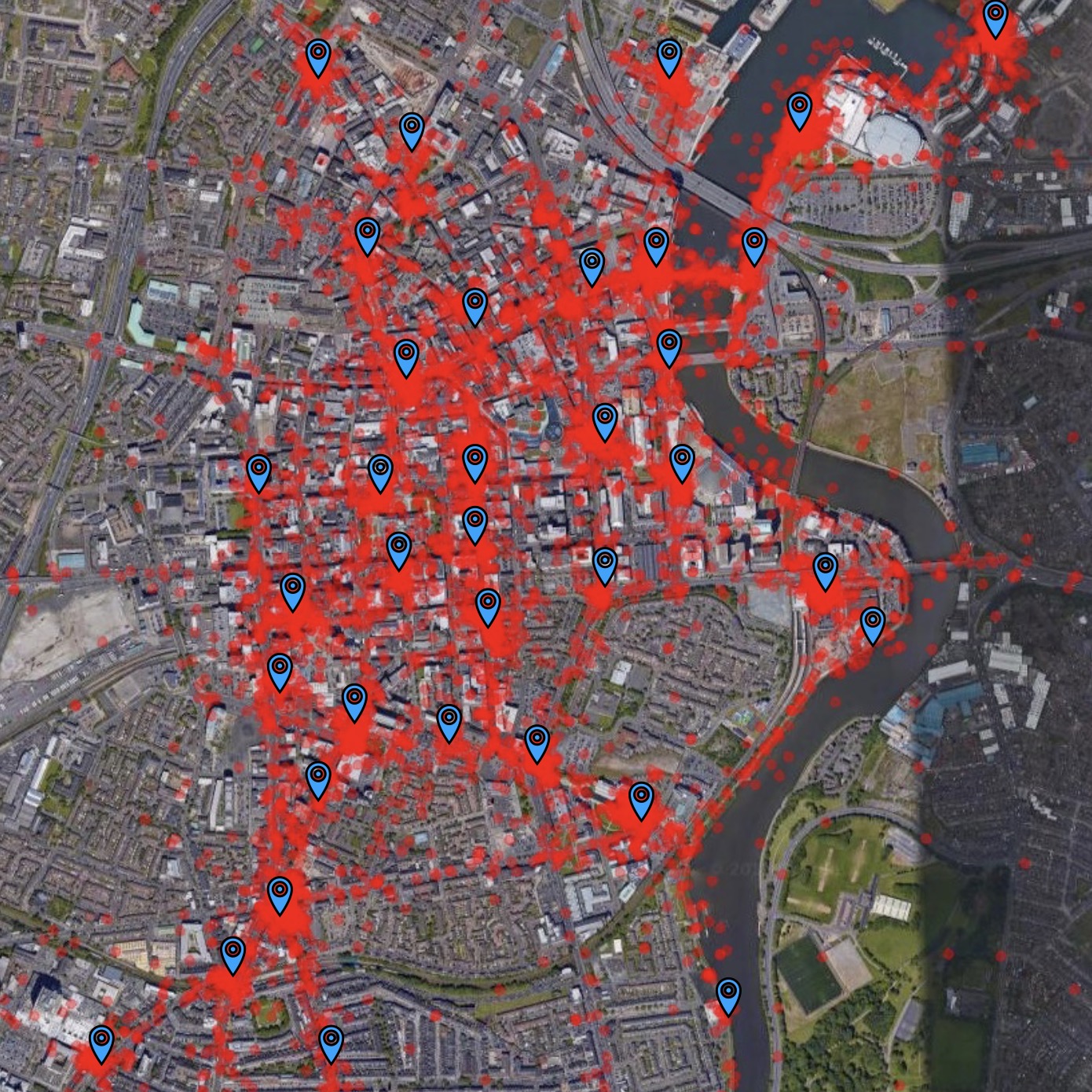}}
		\centerline{(a) urban dense areas}
	\end{minipage}
	\hfill
	\begin{minipage}{0.3\linewidth}
		\centerline{\includegraphics[width=2in]{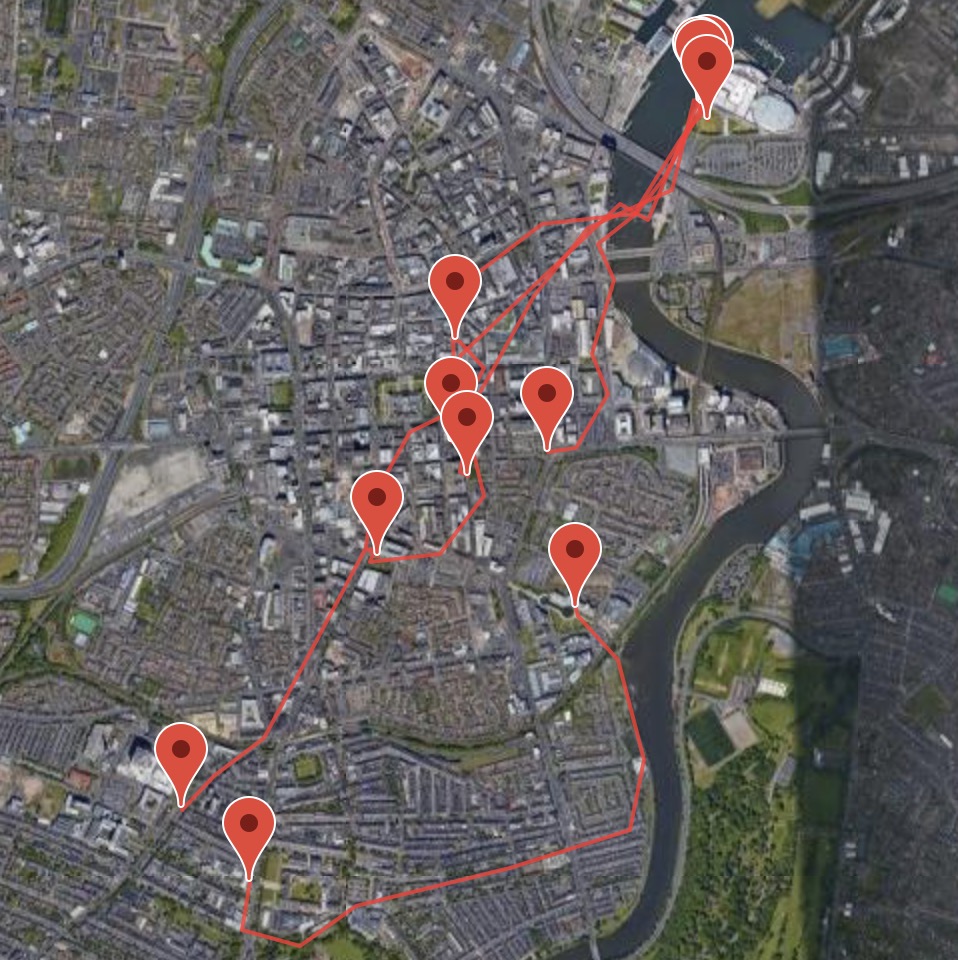}}
		\centerline{(b) trajectory generation}
	\end{minipage}
	\hfill
	\begin{minipage}{0.3\linewidth}
		\centerline{\includegraphics[width=2in]{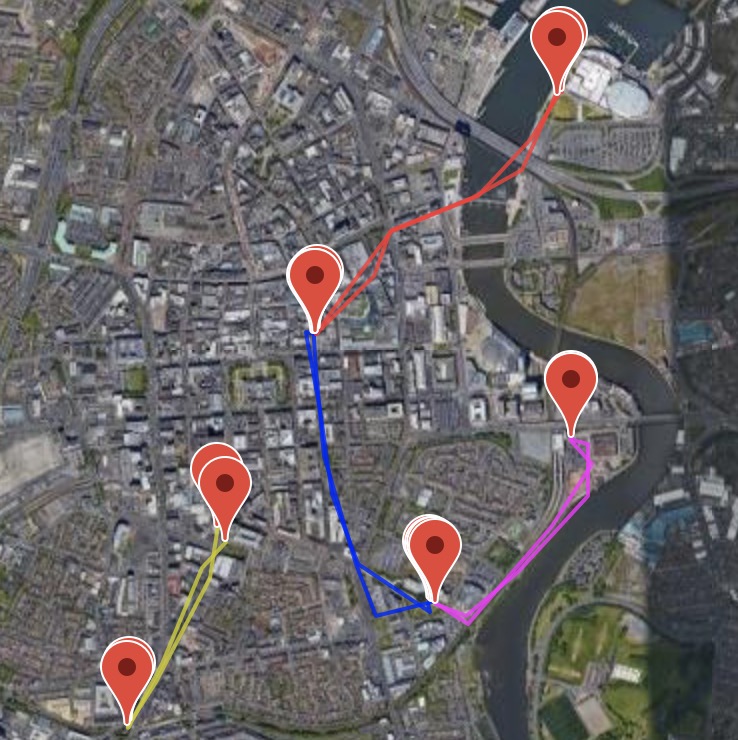}}
		\centerline{(c) similar trajectories}
	\end{minipage}
	\caption{Demonstration of the system}
	\label{fig8}
\end{figure*}
In this demonstration, we used the traffic movement dataset collected for research. Using the Google map API, we mainly display three functions on the map: urban dense area mining, trajectory identification and automatic generation, similar trajectory extraction. Firstly, users need to import GPS data files, and the GPS data contains ID, timestamp, coordinates. The TPM system uses k-means to cluster the GPS data to obtain urban dense areas. Then, after analyzing the trajectory data multiple times, the TPM system gets an appropriate trajectory interval time and automatically generates trajectories. In this step, we have to find an appropriate time interval $\theta$. When the value of $\theta$ changes, the count of trajectory does not change significantly, and that is the appropriate $\theta$. Finally, the system extracts similar trajectories by similarity matching rule. The similarity matching rule is mainly based on the position similarity and sequence of the trajectory. For the similarity of positions, we determine according to the method of clustering. For sequence, we process the trajectory data based on the characteristics of the time series. In addition, we limit the distance of the trajectory to make it close.

Figure \ref{fig8}(a) shows the urban dense areas, where the blue mark is the cluster center, and most of these dense areas are close to Points of Interest (POI), such as subways, schools, restaurants, entertainment venues, etc. Figure \ref{fig8}(b) shows the results of a small number of trajectories in the dataset. Figure \ref{fig8}(c) shows four groups of similar trajectories in different colors. 

\section{Conclusion}
In this paper, we present a GPS-based trajectory pattern mining system called TPM. Firstly, we use K-means++ for clustering the spatial-temporal data, and urban dense areas can be mined. Then, we find a way to get an appropriate time interval of trajectory, and the system automatically generates trajectories after the timing trajectory identification. Mainly, we propose a method for trajectory similarity matching. In terms of application scenarios, the system can be applied to the trajectory system equipped with the GPS device, such as the vehicle trajectory, the bicycle trajectory, the electronic bracelet trajectory, etc., to provide services for traffic navigation and journey recommendation. Meantime, the system can provide support in the decision for urban resource allocation, urban functional region identification, traffic congestion and so on. And this is a trajectory pattern mining system that does not depend on the road network.

\end{document}